\documentclass[a4paper,fleqn,usenatbib]{mnras}

\usepackage{mathptmx}

\usepackage[T1]{fontenc}
\usepackage{ae,aecompl}

\usepackage{graphicx}	

  \title[He abundances and MS colour spread]{On the determination of the He abundance distribution in globular clusters from the width of the main sequence}  
  \author[S. Cassisi et al.] {
  Santi Cassisi,$^{1,2}$\thanks{E-mail:cassisi@oa-teramo.inaf.it (CS)}
  Maurizio Salaris,$^{3}$ 
  Adriano Pietrinferni,$^{1}$ 
  and David Hyder$^{3}$
\\
$^1$INAF~$-$~Osservatorio Astronomico di Teramo, Via M. Maggini, I$-$64100 Teramo, Italy\\
$^2$Instituto de Astrof{\'i}sica de Canarias, Calle Via Lactea s/n, E-38205 La Laguna, Tenerife, Spain\\
$^3$ Astrophysics Research Institute,  Liverpool John Moores University, IC2, Liverpool Science Park, 146 Brownlow Hill, Liverpool L3 5RF, UK
}
\date{Accepted XXX. Received YYY; in original form ZZZ}

\pubyear{2016}

\begin{document}
\label{firstpage}
\pagerange{\pageref{firstpage}--\pageref{lastpage}}
\maketitle

 \begin{abstract}
One crucial piece of information to study the origin of multiple stellar populations in globular clusters, 
is the range of initial helium abundances $\Delta{Y}$ amongst the sub-populations hosted by each 
cluster. These estimates are commonly obtained by measuring the width in colour 
of the unevolved main sequence in an optical colour-magnitude-diagram. 
The measured colour spread is then compared with predictions from theoretical stellar isochrones with varying initial 
He abundances, to determine $\Delta{Y}$. 
The availability of UV/optical magnitudes thanks to the {\sl HST UV Legacy Survey of Galactic GCs} 
project, will allow the homogeneous determination of $\Delta{Y}$ for a large Galactic globular cluster sample. 
From a theoretical point of view, 
accurate UV CMDs can efficiently disentangle the various sub-populations, and main sequence colour differences in the ACS 
$F606W-(F606W-F814W)$ diagram allow an estimate of $\Delta{Y}$. 
We demonstrate that from a theoretical perspective the ($F606W-F814W$) colour is an extremely reliable He-abundance 
indicator. The derivative d$Y$/d($F606W-F814W$), computed at a fixed luminosity along the unevolved main sequence, is 
largely insensitive to the physical assumptions made in stellar model computations, being more sensitive to the 
choice of the bolometric correction scale, and is only slightly dependent on the adopted set of stellar models. From 
a theoretical point of view the ($F606W-F814W$) colour width of the cluster main sequence is therefore a robust diagnostic of 
the $\Delta{Y}$ range.
 \end{abstract}



\section{Introduction}
\label{intro}

During the last fifteen years, our understanding of Galactic globular clusters (GGCs) has been revolutionized 
by a {\it plethora} of observational evidence, based on both photometric and spectroscopic
surveys. The long-standing paradigm that considered these stellar systems
as prototypes of simple (single age, single initial chemical composition) stellar populations
is no longer valid. Indeed GGCs (and old extragalactic globular clusters)
host a {\sl first population} with initial chemical abundance ratios similar
to the one of field halo stars, and additional 
distinct sub-populations, each one characterised by its
own specific variations of the abundances of He, C, N, O, Na, and sometimes
Mg and Al \citep[see, e.g.,][]{gsc}, compared to the field halo abundance ratios.

These abundance patterns give origin, within individual clusters, to well defined (anti-)correlations
between pairs of light elements, the most characteristic one 
being the Na-O anti-correlation, nowadays considered the most prominent signature for the presence of multiple populations
in a given GGC.

According to the currently most debated scenarios 
the observed light element variations are believed to be produced by high temperature proton captures either at the bottom of the
convective envelope of massive asymptotic giant branch (AGB) stars \citep[see, e.g.,][]{agb}, or in the cores of 
main sequence (MS) fast rotating massive stars \citep[FRMSs -- see, e.g.,][]{frms},   
or supermassive stars \citep[SMSs -- see][]{dh2014, d2015}, belonging to the cluster first population.
The CNO processed matter is then transported to the surface either by convection (in AGB stars and the fully convective SMSs)
or rotational mixing (in FRMSs), and injected in the intracluster medium by stellar winds. 
According to these scenarios, the other cluster sub-populations are 
generally envisaged to have formed out of this gas, with a time delay (of the order of at most $10^8$~yr) 
that depends on the type of polluter. Age delays so small, if present, 
are basically impossible to detect from the clusters' colour-magnitude-diagrams (CMDs).

According to the AGB scenario, the observed variation of He amongst stars in the same cluster  
is due to the CNO processed ejecta that are also enriched in He due mainly to the effect of the second dredge up
during the early AGB phase. According to the FRMS and SMS scenario the increase of He comes from the same core H-burning 
that produces the light element variations. Most importantly,
AGB, FRMS and SMS scenarios predict different relationships between light element and He abundance variations.
Accurate estimates of
the He abundance range within individual GGCs is therefore crucial 
to constrain the origin of the multiple population phenomenon 
\citep[see, e.g.,][]{bastian:15,tenorio:16}.

He abundance variations (hereafter $\Delta Y$, where as usual $Y$ denotes the
helium mass fraction) can in principle be determined by direct spectroscopic measurements in
bright, low gravity red giant branch (RGB) stars, and hot horizontal branch (HB) stars.
In case of the RGB stars, $\Delta Y$ estimates (obtained so far for NGC2808 and $\omega$~Centauri)
are based on measurements of the 
near-infrared He I 10830 {\AA}~chromospheric line \citep[see][]{dupree:11,pasq,da}.
Uncertainties related to the necessary non-LTE corrections and chromospheric modelling somewhat limit
the current accuracy of this diagnostic.

Spectroscopic measurements of the He abundance in HB stars \citep[see, e.g.,][and references therein]{vgpg,marino:14,mll} 
are based on several photospheric He lines that are excited
in HB stars hotter than $\sim$8500~K, but only objects between this lower $T_{eff}$ limit, and an upper limit
of about 11500-12000~K are appropriate targets. Above this $T_{eff}$ threshold the gravitational settling of He sets in
\citep[see, e.g.,][]{mb,moehler:12,moehler:14},
and the measured abundances would be much lower than the initial values\footnote{In case of both RGB and HB spectroscopy,
the measured He abundances are equal to the initial values, plus the small increase 
due to the first dredge up, that can be 
easily predicted by stellar models. In case of efficient atomic diffusion during the MS, the RGB surface He 
abundances after the first dredge up are slightly lower than the initial ones \citep{pv}}.
These requirements about the $T_{eff}$ of the targets hamper the use of this method to estimate reliably $\Delta Y$ ranges, 
for the following reasons. First of all, not all GGCs host HB stars that reach $T_{eff}$=8500~K 
(a typical example is the metal rich cluster 47~Tucanae). Secondly, in GGCs with an 
extended HB the various sub-populations show some kind of colour segregation along the branch.
The first population with primordial He abundance is preferentially located at the red side of
the HB, whilst sub-populations with increasing  degree of variations of the light element abundances
(hence presumably He) are distributed towards the hotter end of the HB \citep[see, e.g.,][]{marino:11,gls}.
This implies that the $T_{eff}$ window suitable to determine $Y$ in HB stars cannot sample the full  
$\Delta Y$ range in the cluster.

The presence of He enhanced stellar populations affects also the CMD 
of GGCs, and indeed, the existence of a range of initial He abundances within individual GGCs 
was first {\bf inferred} from the optical CMD of MS stars in $\omega$~Centauri \citep[][]{bedin:04,norris:04,kingomega} and NGC~2808 \citep[][]{p07}. 
The details of the techniques applied to determine $\Delta Y$ ranges from CMDs vary from author to author  
\citep[compare for example the methods applied by][]{kingomega, milone:12,milone:15}, but they are all based on the following 
properties of stellar models/isochrones.

It is well established that an increase of the initial He abundance in MS stars of a given mass changes their   
$T_{eff}$ as a consequence of the associated decrease of the envelope radiative opacities, so that they 
become hotter with increasing $Y$. At the same time an increase of $Y$ increases the
molecular weight of the stellar gas and makes MS stars of a given mass also brighter \citep[see, e.g.,][]{kww,cs13}.
The net effect is that, at a given brightness, the unevolved MS of theoretical isochrones with increasing He abundances (and fixed metal content)
become progressively bluer.

In addition, \citet{swff}, \citet{sbordone} and \citet{cassisi:13} have shown that optical CMDs of GGCs
are expected to display multiple MS sequences only if the different
populations are characterized by variations of $Y$. 
The reason is that for a given $Y$, {\sl standard} $\alpha$-enhanced isochrones (from the MS to the tip of the RGB) 
that match the metal composition of the cluster first population, 
are identical to isochrones that include the full range of observed light element variations. 
Moreover, the same light element abundance variations (and variations of $Y$)
do not affect the bolometric corrections to optical photometric filters \citep[see][]{gcbn, sbordone}.

To summarize, it is possible to safely determine the range $\Delta Y$ in individual GGCs by comparing theoretical MS isochrones 
with standard $\alpha$-enhanced metal distributions and varying initial $Y$, to cluster optical CMDs.
Once photometric errors (and eventually differential reddening and unresolved binaries) are taken into account,
the observed width of the unevolved MS provides an estimate of $\Delta Y$ amongst the cluster multiple populations.

This method can be applied to all clusters with available accurate optical photometry provided that the various 
sub-populations have been identified (as discussed below). The colour separation of their respective MSs is an especially reliable 
$\Delta Y$ diagnostic, based on a {\sl simple} and well modelled evolutionary phase.
Other methods based on cluster photometry,
such as comparisons of the predicted and observed brightness of the luminosity function RGB bump \citep[see, e.g.,][]{milone:15}, 
and the analysis of the HB 
morphology and brightness with synthetic HB models \citep[see,e.g.,][]{dale, gls} rely on models of more advanced evolutionary phases, and potentially
subject to larger systematic errors.

In this paper we study the robustness of the theoretical calibration of MS colour versus initial $Y$ in optical filters,
in view of its relevance for the quest of the origin of the multiple populations in globular clusters, and the availability 
of a large sample of high-precision UV-optical CMDs from the {\sl Hubble Space Telescope UV Legacy Survey of Galactic GCs}. 
This survey has observed 54 GGCs through the filters $F275W$, $F336W$, $F438W$ of the Wide Field Camera 3 (WFC3) 
on board $HST$ \citep[][]{uvhst}, and its results {\bf must be used} in conjunction with the existing $F606W$ and $F814W$ 
extensive and homogeneous photometry from 
the $HST$/ACS GGC Treasury project \citep[see][]{acs} to identify the various sub-populations hosted by individual clusters  
\citep[see, e.g.,][and references therein]{nardiello, milone:15} and determine $\Delta Y$ homogeneously, for a large GGC sample.
The use of UV filters maximizes the colour separation between the various populations in individual clusters (because the 
CNONa variations affect the bolometric corrections in UV filters), and after the different MS components are clearly 
identified in the UV, their (smaller) colour separation in optical filters provides $\Delta Y$ on a cluster-by-cluster basis.

\citet{bastian:15} have shown how none of the proposed AGB, FRMS or SMS based scenarios are able to match simultaneously 
$\Delta Y$ and the observed range of the O-Na anticorrelation in a small sample of GGCs where $\Delta Y$ estimates are available, 
questioning the reliability of our current ideas about the formation of cluster multiple populations. An homogeneous 
determination of $\Delta Y$ for a large sample of GGCs is strongly required to confirm or disprove this result, and 
this will be done within the {\sl Hubble Space Telescope UV Legacy Survey of Galactic GCs} project. 
Our analysis aims at assessing the sources of theoretical systematic errors involved in the determination of $\Delta Y$ 
from MS photometry in optical $HST$-ACS CMDs, using theoretical stellar isochrones. 
We focus the analysis that follows on the ACS $F606W-(F606W-F814W)$ CMD (that roughly corresponds 
to the Johnson-Cousins $VI$ CMD), because these are the only filters available within the survey,  
that are completely unaffected by the presence of CNONa abundance variations.
The UV filters but also the WFC3 $F438W$ filter (or the $F435W$ ACS counterpart) are sensitive 
--albeit to different degrees-- to these abundance variations \citep[see, e.g.,][]{uvhst}. 
In particular, using the theoretical spectra by \citet{sbordone} for [Fe/H]=$-$1.6 for an unevolved MS star 
along a 12~Gyr isochrone (with $T_{eff}$=4621 and surface gravity log($g$)=4.77, both for a first population standard 
$\alpha$-enhanced mixture, and composition with CNONa anticorrelations at constant CNO sum) we found a difference    
of about 0.05~mag between the corresponding bolometric corrections for the WFC3 $F438W$ filter (and the ACS $F435W$ counterpart), whereas 
the same differences are $\sim$0.001~mag for both the $F606W$ and $F814W$ ACS filters. This confirms that the observed intrinsic 
colour range of the MS in the ACS $F606W-(F606W-F814W)$ CMD is due to $Y$ differences, 
irrespective of the exact values of the C, N, O, Na abundances.  

The plan of the paper is as follows:
Section~\ref{models} presents the reference evolutionary framework and the additional isochrone sets 
computed to study the effect of various physics inputs on the derived $\Delta Y$; Section~\ref{analysis} compares 
the various isochrones sets to establish the robustness of the MS d$Y$/d(colour) derivatives. 
A critical summary of the results will close the paper.

\section{The theoretical models}
\label{models}

The stellar models and isochrones employed in our analysis have been computed with the
BaSTI stellar evolution code, and 
our {\sl reference} calculations employ the input physics 
and bolometric corrections fully described in \citet{basti} and \citet{bastia}.
We have explored a large metallicity range, from [Fe/H]$=-$2.1 to [Fe/H]=$-$0.7 
([$\alpha$/Fe]=0.4), that covers most of the metallicity range spanned by the GGC population.
In the following we detail the case with [Fe/H]=$-$1.3, representative of the results that we 
obtain at all metallicities.

More in detail, we calculated models and isochrones (neglecting the effect of atomic diffusion) with metal mass fraction $Z$=0.002, $Y$=0.248 and 
[$\alpha$/Fe]=0.4, corresponding to [Fe/H]=$-$1.3, that represent the first population in a cluster with this [Fe/H].
In addition we calculated
models and isochrones for the same $Z$ and $\alpha$-enhancement, but with $Y$=0.30, 0.35 and 0.40, respectively,
corresponding to sub-populations with increasing initial He abundance\footnote{All these stellar models and isochrones are available at the BaSTI
URL site: http://www.oa-teramo.inaf.it/BASTI}.

\begin{figure}
\centering
\includegraphics[width=\columnwidth]{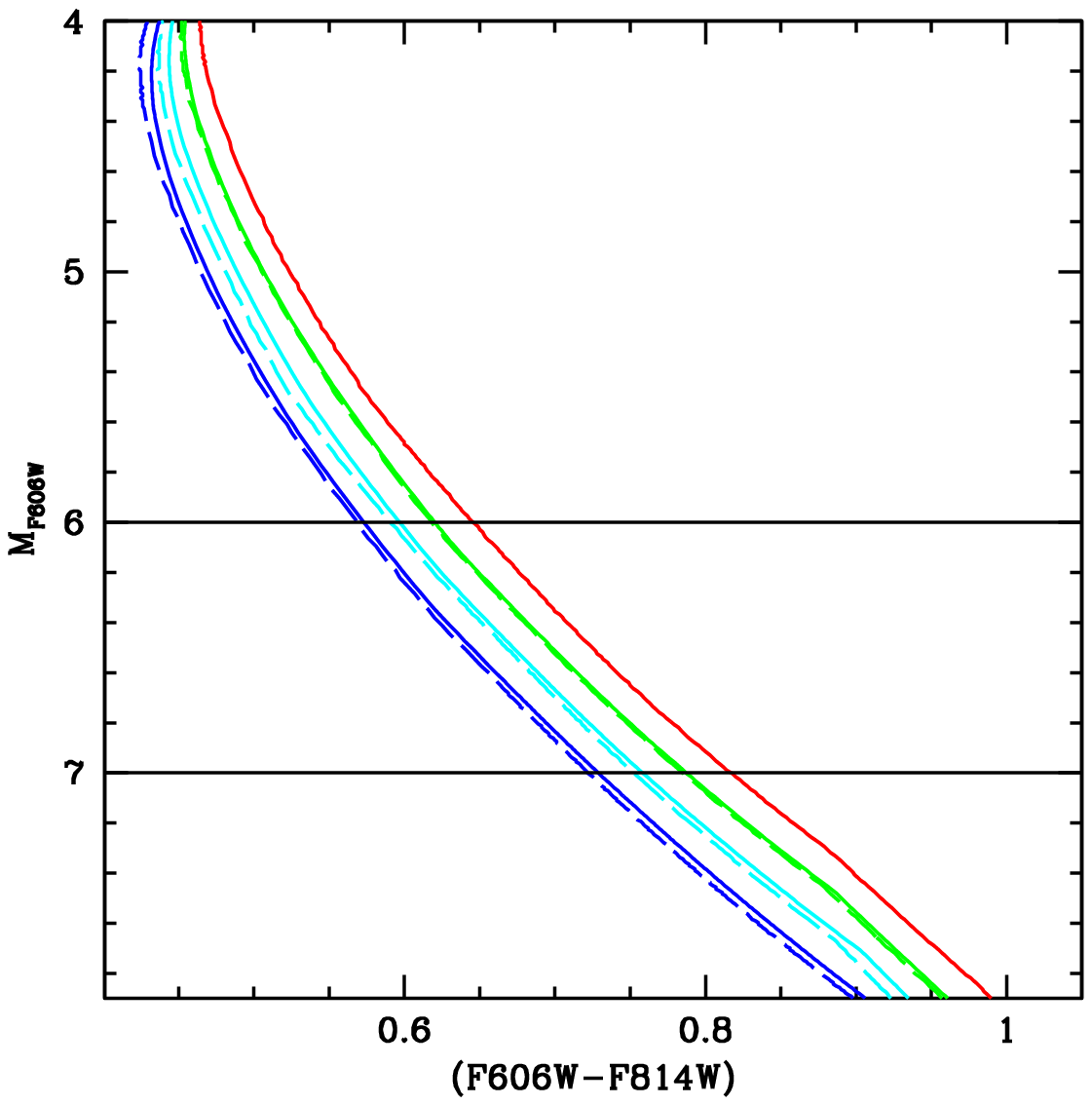}
\caption{A set of our reference MS isochrones (calculated at fixed $Z$) for an age of 12~Gyr and (moving towards bluer colours) 
$Y$=0.248, 0.30, 0.35, 0.40, respectively (solid lines), compared to MS isochrones calculated with the same 
age and $Y$ values, but keeping [Fe/H] constant (dashed lines -- see text for details).}
\label{isoref}
\end{figure}

It is important to clarify the choice of varying $Y$ in the models while keeping $Z$ fixed, instead of keeping [Fe/H] fixed. 
The observed abundance patterns do not change the total metallicity $Z$. This is essentially because 
the sum of the CNO abundances plus the abundances of all elements not affected by the anticorrelations 
(chiefly amongst them Ne, Fe and Si) make about 90\% or more of the metal content by mass, and 
are unchanged amongst the cluster sub-populations, with the exception of a few  cases like NGC1851 
\citep{yong:15}, NGC6656 \citep{marino:12a}, and $\omega$~Centauri \citep{marino:12b}.
At the same time, as shown by \citet{brag}, when the range of $Y$ amongst the cluster sub-populations 
is sufficiently large, it is possible to detect corresponding small variations of [Fe/H] consistent with 
the expected change due to the variation of the hydrogen 
abundance (notice that $\Delta Y$=0.1 at fixed Fe corresponds to $\Delta$[Fe/H]$\sim$0.06). 

As already stated, we focus the analysis that follows on the $F606W-(F606W-F814W)$ CMD, 
based on $HST$-ACS filters.
As mentioned in the introduction, light element variations do not affect bolometric corrections 
to these filters, and moreover the $(F606W-F814W)-T_{eff}$ relation is very weakly dependent on metallicity, as we have 
also verified with our adopted reference bolometric corrections based on ATLAS9 model atmospheres 
and spectra \citep[see][]{basti}. This means that the predicted MS colour shifts in this CMD 
are more closely related to the $T_{eff}$ shifts caused by $Y$ variations, and also that scaled solar 
BCs, with either the same total metallicity or the same [Fe/H] of the $\alpha$-enhanced composition, can be safely used 
(if the appropriate $\alpha$-enhanced BCs are unavailable) as demonstrated in Fig.~1 of \citet{transf}.
The BCs applied to our isochrones have been calculated for the [Fe/H] of the 
{\sl normal} $Y$ composition (the lowest $Y$ value) and kept unchanged for the He-enhanced isochrones, 
because they are unaffected by a change of $Y$ when the total metal content is kept constant 
(and they are unaffected by the CNONa anticorrelations -- see Sect.~\ref{intro}).


Figure~\ref{isoref} displays the MS of a set of our {\sl reference} isochrones, for an age of 12~Gyr and
varying $Y$. The horizontal lines mark $M_{F606W}$=6 and 7, respectively, the two magnitude levels along 
the unevolved MS --more than $\sim$2~mag below the turn-off level-- 
where we will calculate the derivative d$Y$/d($F606W-F814W$) and discuss its uncertainties
\footnote{We have obtained quantitatively the same fractional uncertainties on d$Y$/d($F606W-F814W$) when 
we considered as reference magnitudes $M_{F814W}$=5.6 and 6.6, corresponding to $\sim$2 and $\sim$3 
magnitudes below the MS turn off in the $F814W$ filter}. The same figure shows, for the sake of comparison, isochrones
calculated for the same age and initial $Y$ values, but keeping [Fe/H]=$-$1.3 for all $Y$ 
--hence $Z$ decreases with increasing $Y$, by at most $\sim$15\%.
For $\Delta Y$ up to $\sim$0.05, there is essentially no difference between the case of keeping $Z$ or [Fe/H] fixed.
Differences appear when $\Delta Y>$0.05 and increase with increasing $Y$, due to the fact that $Z$ must be 
decreased by increasingly larger values in order to keep [Fe/H] unchanged. As a result, employing theoretical MSs 
at varying $Y$ and [Fe/H] fixed, would lead to systematically smaller $\Delta Y$ values, when $\Delta Y>$0.05.

For our reference isochrones, 
at $M_{F606W}$=7 the evolving mass along the $Y$=0.248 MS is equal to 0.57$M_{\odot}$, decreasing to 
0.46$M_{\odot}$ along the $Y$=0.40 MS. At $M_{F606W}$=6 the evolving masses are equal to 0.65$M_{\odot}$ and 
0.52$M_{\odot}$, respectively.
The MSs with higher initial $Y$ are bluer at a given $M_{F606W}$, and the colour differences between 
pairs of MSs with different initial $Y$ tend to slowly decrease moving to brighter magnitudes. 

From these isochrones we obtain d$Y$/d($F606W-F814W$)=1.8  
at $M_{F606W}$=7 and d$Y$/d($F606W-F814W$)=2.1 at $M_{F606W}$=6, irrespective of the value of $Y$, within the range 
$Y$=0.248-0.40. These derived  values of d$Y$/d($F606W-F814W$) are, as expected, 
insensitive to the adopted isochrone age $t$. We have actually verified that variations 
of $t$ by $\pm$1-2~Gyr around $t$=12~Gyr do not affect d$Y$/d($F606W-F814W$) at both  $M_{F606W}$=6 and 7.

We have then calculated additional isochrones for the same $Z$ and the same $Y$ range, 
but varying one at a time the following inputs, that affect the CMD location of evolutionary tracks and isochrones:

\begin{enumerate}
\item{The BCs adopted to compare theoretical isochrones with observed CMDs;}
\item{the efficiency of convection in the outer, superadiabatic layers, e.g. the value of the mixing length parameter $\alpha_{\rm MLT}$ employed in the model computations;}
\item{the outer boundary conditions employed in the stellar model calculations;}
\item{inclusion of fully efficient atomic diffusion everywhere in the models, or just below the convective envelope.}  
\end{enumerate}

With these model computations we have calculated appropriate sets of isochrones, to assess the 
robustness of the d$Y$/d($F606W-F814$) predictions, that will be discussed in the next section.

Finally, to explore the effect of using indepedent evolutionary calculations --that employ also some different choices of physics inputs-- 
we considered the \citet{pisa} and \citet{vdbmodels} models, that provide isochrones at fixed $Z$ and varying $Y$ values for several ages.

\section{Analysis}
\label{analysis}

For all additional sets of isochrones listed in the previous section we have determined 
d$Y$/d($F606W-F814W$) at both $M_{F606W}$=6 and 7, and compared these values with the results from the reference 
set. The results of these comparisons are discussed in the following sections, 
considering separately each one of the new sets.

\subsection{Bolometric corrections}

To study the effect of varying the bolometric corrections BCs, 
we employed the theoretical values calculated with the PHOENIX and MARCS model atmosphere codes   
\citep[see][]{bh, cv} respectively, and the empirical BCs by \citet{wl}.  
\citet{wl} provide BCs to the Johnson-Cousins $VI$ filters, that we have translated to the 
corresponding ACS filters using the empirical relationships by 
\citet{sirianni}.
Both PHOENIX and MARCS based BCs are calculated for an $\alpha$-enhanced [$\alpha$/Fe]=0.4 
metal mixture; however the reference solar heavy element distribution for the available MARCS models is from \citet{gas}, instead of \citet{gn} 
as for PHOENIX and our reference ATLAS9 BCs. This difference may possibly add a degree of inconsistency to the results of this comparison.

\begin{figure}
\centering
\includegraphics[width=\columnwidth]{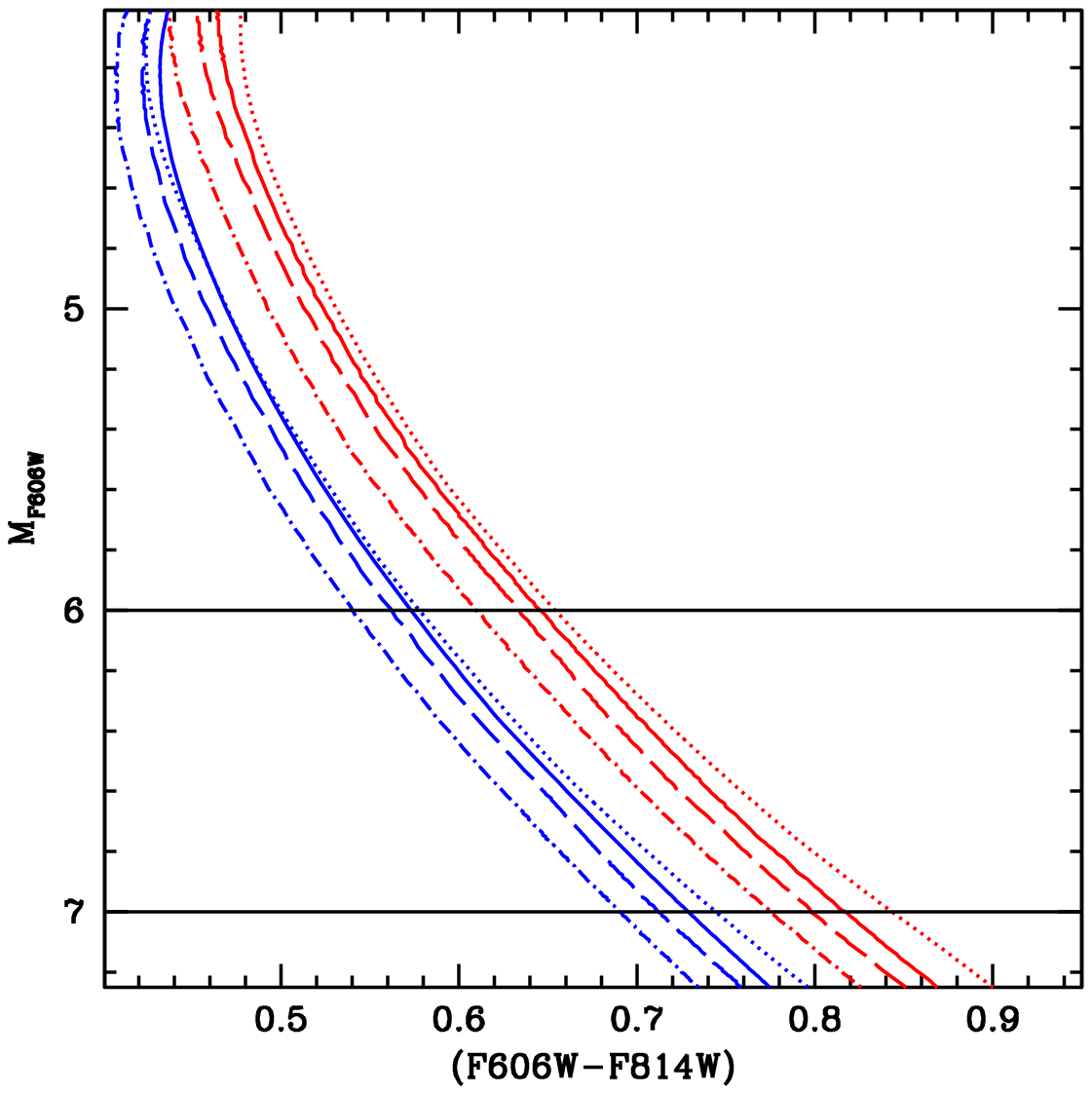}
\caption{Comparison of 12~Gyr MS isochrones for $Y$=0.248 and 0.40. We display our reference 
isochrones (with ATLAS9 bolometric corrections, solid lines), and the same isochrones but applying alternative 
empirical (dotted lines) and theoretical (PHOENIX as dashed lines, MARCS as dash-dotted lines) 
bolometric corrections, respectively (see text for details).}
\label{isobcs}
\end{figure}

Figure~\ref{isobcs} compares these new sets of isochrones with the reference ones, for $Y$=0.248 and 0.40 (the extreme 
values of the range spanned by our calculations), $t$=12~Gyr.
The PHOENIX and MARCS BCs produce MSs bluer by about 0.015 and 0.04~mag respectively, at all $M_{F606W}$, compared to the reference set.
The shift is the same for both values of $Y$, and as a consequence we find that d$Y$/d($F606W-F814W$) stays unchanged 
compared to the reference values at $M_{F606W}$=6 and 7.

The results employing the empirical BCs by \citet{wl} are different.
In this case it is evident from Fig.~\ref{isobcs} that there is a shift in the MS colours (generally redder than 
the reference isochrones) but also a clear change of shape. We obtain 
d$Y$/d($F606W-F814W$)=2.0 and 1.5 -- irrespective if $Y$ -- at  $M_{F606W}$=6 and 7, respectively. The difference with respect 
to the reference derivatives are larger at the fainter magnitude level.

\subsection{The value of $\alpha_{\rm MLT}$ and the outer boundary conditions}

The mixing length theory \citep[MLT --][]{bv} is almost universally used to compute the temperature gradient in 
superadiabatic layers of stellar (interior and atmosphere) models. 
This formalism contains in its standard form 4 free parameters. Three parameters 
are fixed a priori (and define what we denote as the MLT {\sl flavour}) 
whereas one (the so-called mixing length, $\alpha_{\rm MLT}$) is calibrated by 
reproducing observational constraints \citep[see, e.g.,][and references therein]{mlt}.
In our reference calculations we have employed the \citet{cg} implementation of the MLT 
\citep[corresponding to the ML1 {\sl flavour} described in][]{mlt} with $\alpha_{\rm MLT}$ 
fixed by the standard solar model calibration \citep[see][for details]{basti}.


In these additional sets of isochrones we have considered a variation of $\alpha_{\rm MLT}$ by $\pm$0.1 around the solar 
value. This is consistent with the variation of $\alpha_{\rm MLT}$ along the isochrone MS predicted by the 3D radiation hydrodynamics 
simulations by \citet{mwa}.
We find that at fixed $Y$ the three sets of MSs 
are virtually coincident for $M_{F606W}\sim$6 and fainter magnitudes. The reason is that models in this magnitude range 
have deep convective envelopes almost completely adiabatic, and the variation of $\alpha_{\rm MLT}$ has a very minor effect 
on their temperature stratification. As a result, d$Y$/d($F606W-F814W$) values are unchanged.

Given that the envelopes of stellar models in this magnitude range are almost completely adiabatic, we expect a negligible 
difference when employing alternatively the \citet{cm} theory of superadiabatic convection, instead of the MLT. This is confirmed by 
Fig.~A2 in \citet{mdc}, that compares in the $L$-$T_{eff}$ diagram GGC isochrones calculated with a solar calibrated mixing length 
and with the \citet{cm} formalism. In the relevant $L$ range (between log($L/L_{\odot})\sim -$0.5 and $\sim -$0.9) the two sets 
of MS isochrones at a given initial chemical composition are virtually indistinguishable.

Regarding the model outer boundary conditions, we recall that 
to integrate the stellar structure equations, it is necessary
to fix the value of the pressure and temperature at the
surface of the star, usually close to the photosphere  \citep[see, e.g.,][and references therein]{don:08}. In our reference calculations 
we have determined this boundary condition by integrating 
the atmospheric layers using the solar \citet{ks} $T(\tau)$ relation, 
supplemented by the hydrostatic equilibrium condition and the equation of state. 
The choice of the outer boundary conditions is important in this context, because 
it affects the $T_{eff}$ of stellar models with convective envelopes \citep[see, e.g.,][for more details]{cs13,sc:15}

In this additional set of MS isochrones we have employed the 
Eddington grey $T(\tau)$ instead. We have first recalibrated the solar value of the 
mixing length $\alpha_{\rm MLT}$ with this $T(\tau)$ relation, and then calculated 
models and isochrones with the new solar calibrated mixing length (about 0.1 smaller 
than the value calibrated with the reference $T(\tau)$ relation).

\begin{figure}
\centering
\includegraphics[width=\columnwidth]{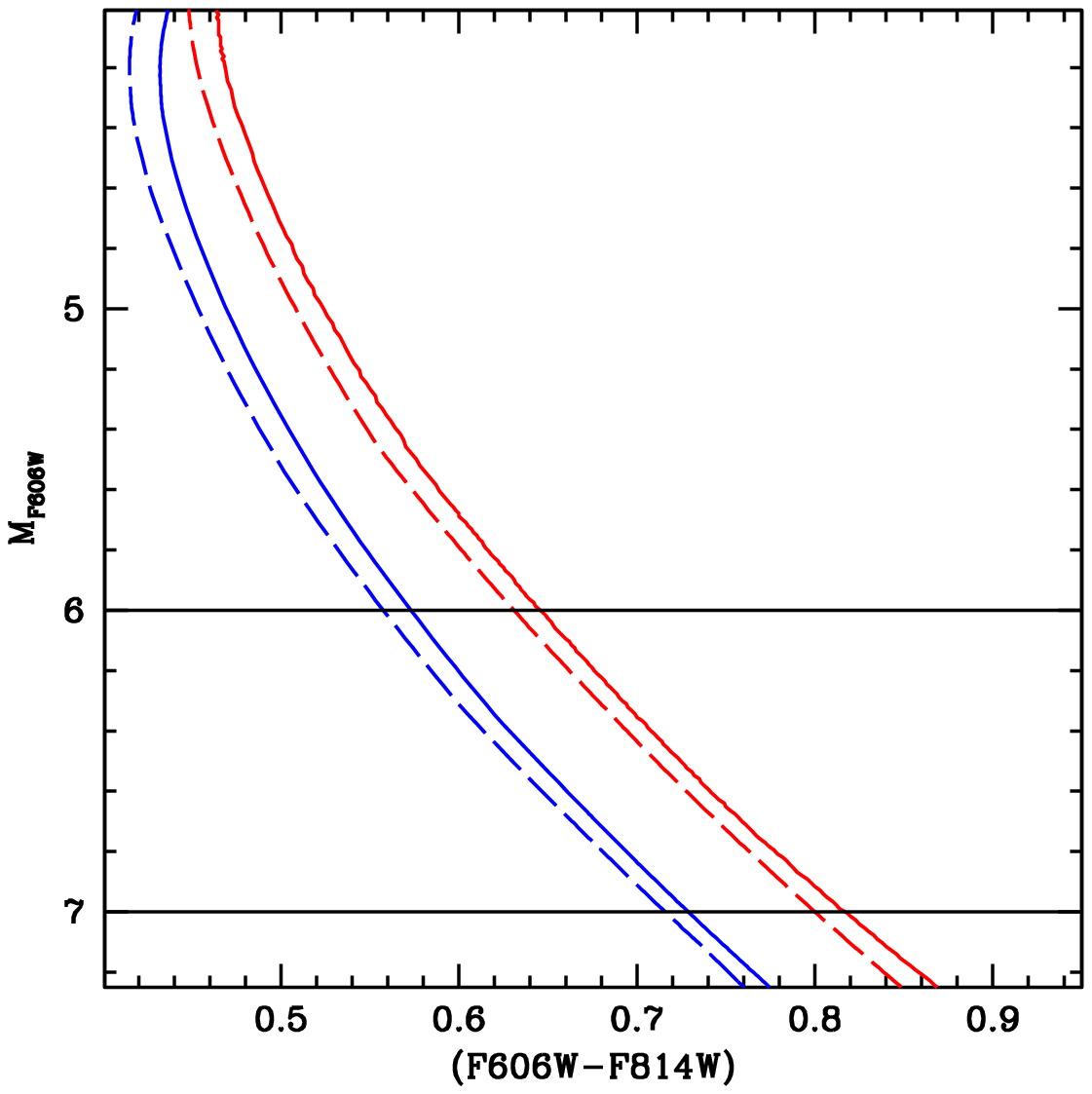}
\caption{As Fig.~\ref{isobcs} but displaying our reference 
isochrones (solid lines), and isochrones calculated using the Eddington grey $T(\tau)$ for 
the outer boundary conditions (dashed lines -- see text for details).}
\label{isobc}
\end{figure}

Figure~\ref{isobc} compares these new calculations with the reference one for the same selected $t$ and $Y$ values 
employed in the previous comparisons.
The MSs calculated with the Eddington $T(\tau)$ are systematically bluer (by about 0.02~mag), but the shift is independent 
of $Y$ and the MS shape is preserved. As a consequence d$Y$/d($F606W-F814W$) values are unchanged compared to the reference case.

\subsection{The efficiency of atomic diffusion}

Asteroseismic data require the inclusion of atomic diffusion in standard solar model calculations, to match 
the inferred solar sound-speed profile, depth of the convective envelope and surface He abundance 
\citep[see, e.g.,][and references therein]{bpw, villante}. 
The situation for GGC stars is however very different. Spectroscopic determinations of metal abundances 
from the MS turn off to the RGB of a number of GGCs, have shown that the efficiency of atomic diffusion from the convective envelope 
of old metal poor stars is severely reduced \citep[see, e.g.,][and references therein]{grnodiff,lind,msl,gruyt} 
by some as yet unspecified competing physical process.
Nothing is known about the efficiency of diffusion in the radiative interiors of these objects.

To study the potential impact on d$Y$/d($F606W-F814W$) 
we have therefore calculated two sets of isochrones. The first one includes fully efficient diffusion throughout the 
stellar models \citep[we employed the diffusion coefficient by][implemented in the BaSTI code]{tbl}, 
and one including efficient atomic diffusion only below the boundary of the convective envelope. 

The set with fully efficient diffusion is an extreme and, for what we know, unrealistic case for GGC stars, but 
necessary to understand the comparisons with d$Y$/d($F606W-F814W$) values obtained with other theoretical isochrones 
that include atomic diffusion (see next section).
Regarding the models with artifically inhibited diffusion from the envelope, 
starting at a distance of 0.5$H_P$ (where $H_p$ is the pressure scale height 
at the Schwarzschild boundary of the convective envelope) below the surface convection boundary, we  
smoothly reduced the diffusion velocities of the various elements (with a quadratic function of radius) 
so that they reached zero at the bottom of the fully mixed envelope. 
This second set is a numerical experiment to simulate the inhibition of diffusion from the convective boundary, 
keeping the element transport efficient in the interiors.

In general, the effect of atomic diffusion on stellar evolution tracks of a given mass (for masses 
typical of GGCs) and initial chemical composition 
is to shorten the MS lifetimes by $\sim$1~Gyr due to the diffusion of helium to the centre, 
that is equivalent to a faster aging of the star. Also, due to the diffusion of He and metals from the convective envelopes, 
the surface H abundance increases, causing a higher opacity and a shift of the tracks towards redder colours. 
In addition, the inward settling of helium raises the core molecular weight and the molecular weight
gradient between surface and centre of the star, that also contributes to increase 
the stellar radius \citep[see, e.g.,][]{ccdw,sgw}.

\begin{figure}
\centering
\includegraphics[width=\columnwidth]{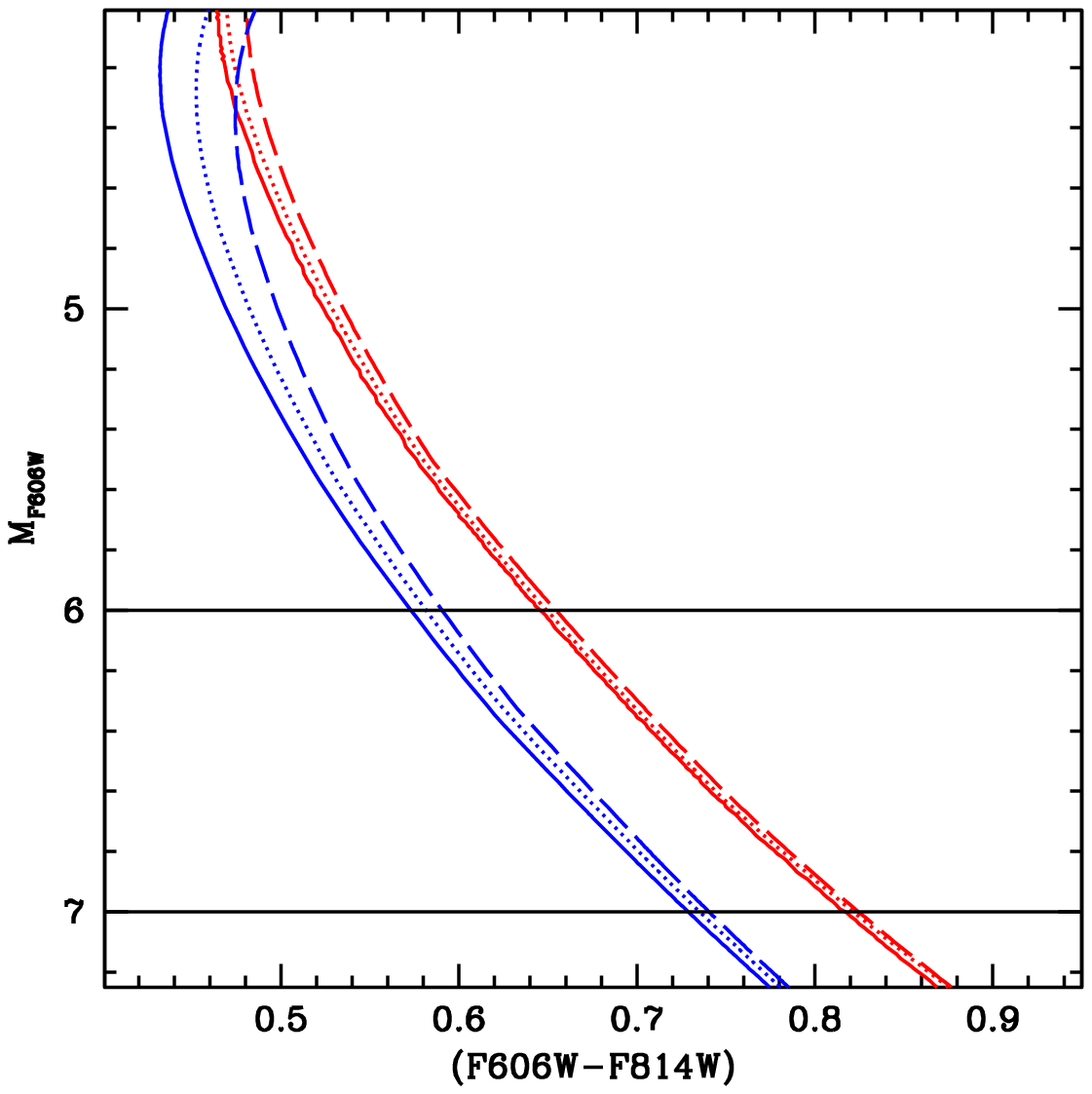}
\caption{As Fig.~\ref{isobcs} but displaying our reference 
isochrones (solid line), and 11~Gyr old isochrones including fully efficient atomic diffusion (dashed lines) 
and diffusion efficient only below the convective envelope (dotted lines - see text for details).}
\label{isodiff}
\end{figure}

Figure~\ref{isodiff} shows the results of these calculations, compared to the reference case. The displayed isochrones with 
diffusion are $\sim1$~Gyr younger than the reference one, so that they have approximately the same MS turn off magnitude. 
As for the case of individual tracks, fully efficient atomic diffusion shifts the 
isochrone MS to redder colours, by an amount that increases moving towards 
the MS turn off. Lower MS objects are less affected because of their very extended, fully mixed, convective envelopes, 
that minimize the reduction of surface metals and He.
The effect is larger when $Y$ increases, because of generally shallower convective envelopes. 

We find that d$Y$/d($F606W-F814W$) at $M_{F606W}$=6 increases compared to the 
reference value, but only for $Y>$0.30, whilst it is basically 
unchanged at $M_{F606W}$=7, where the effect of diffusion is negligible.
When $Y>$0.30, the value of d$Y$/d($F606W-F814W$) in case of fully efficient diffusion depends on the actual 
value of $Y$. We find that for $Y$ between 0.30 and 0.35 d$Y$/d($F606W-F814W$)=2.5, increasing 
to 2.6 when $Y$ is between 0.35 and 0.40.

It is instructive to notice that also the MS with diffusion from the convective layers inhibited is shifted to the red 
compared to reference isochrones, although by a smaller amount compared to the case of fully efficient diffusion. 
Most importantly, the shifts are $Y$- and magnitude-independent, and the effect on d$Y$/d($F606W-F814W$) is negligible.

\subsection{Independent stellar model calculations}

We have also considered two publicly avalaible independent sets of stellar evolution models and isochrones 
that include variations of $Y$ at constant $Z$, to explore the effect of varying additional input physics. 
We considered the \citet{pisa} and \citet{vdbmodels} model databases. 

The \citet{pisa} database provides isochrones in the theoretical $L$-$T_{eff}$ diagram  
at fixed $Z$ and varying $Y$ values for several ages. We focused on their 12~Gyr,  
$Z$=0.002 isochrones with $Y$=0.25, 0.33 and 0.38, respectively. 
They have been calculated for a scaled solar chemical composition, and employ the solar mixture by \citet{agss}, different from our reference solar mixture 
\citep[][]{gn}. These calculations rely also on some physics inputs different from those adopted in our own calculations. More in detail, 
\citet{pisa} adopted the OPAL equation of state \citep[][]{opal}, whilst our reference calculations employ the freeEOS one \citep[see][]{cassisi:03,basti}; 
the outer boundary conditions for MS stars at these 
metallicity and ages are taken from the PHOENIX model atmospheres \citep[][]{bh} 
instead of a $T(\tau)$ integration. Also, the reaction rate for the $^{14}N(p,\gamma)^{15}O$ reaction is taken from 
\citet{imbri}, whilst our reference calculations employ the \citet{nacre} rate.
Finally, these isochrones include also fully efficient atomic diffusion using \citet{tbl} diffusion coefficients.

We applied to these isochrones our scaled solar ATLAS9 BCs calculated for the \citet{gn} 
solar mixture. We have verified beforehand, with calculations performed using the ATLAS9 suite of codes \citep[][]{atlas}, that for 
MS stars in the $T_{eff}$ and $L$ range relevant to this analysis, the  BCs for the $F606W$ and $F814W$ filters are the same 
when employing the \citet{agss} or the \citet{gn} solar heavy element distributions. This means that our reference ATLAS9 BCs 
are also appropriate for the chemical composition adopted by \citet{pisa}. 

\begin{figure}
\centering
\includegraphics[width=\columnwidth]{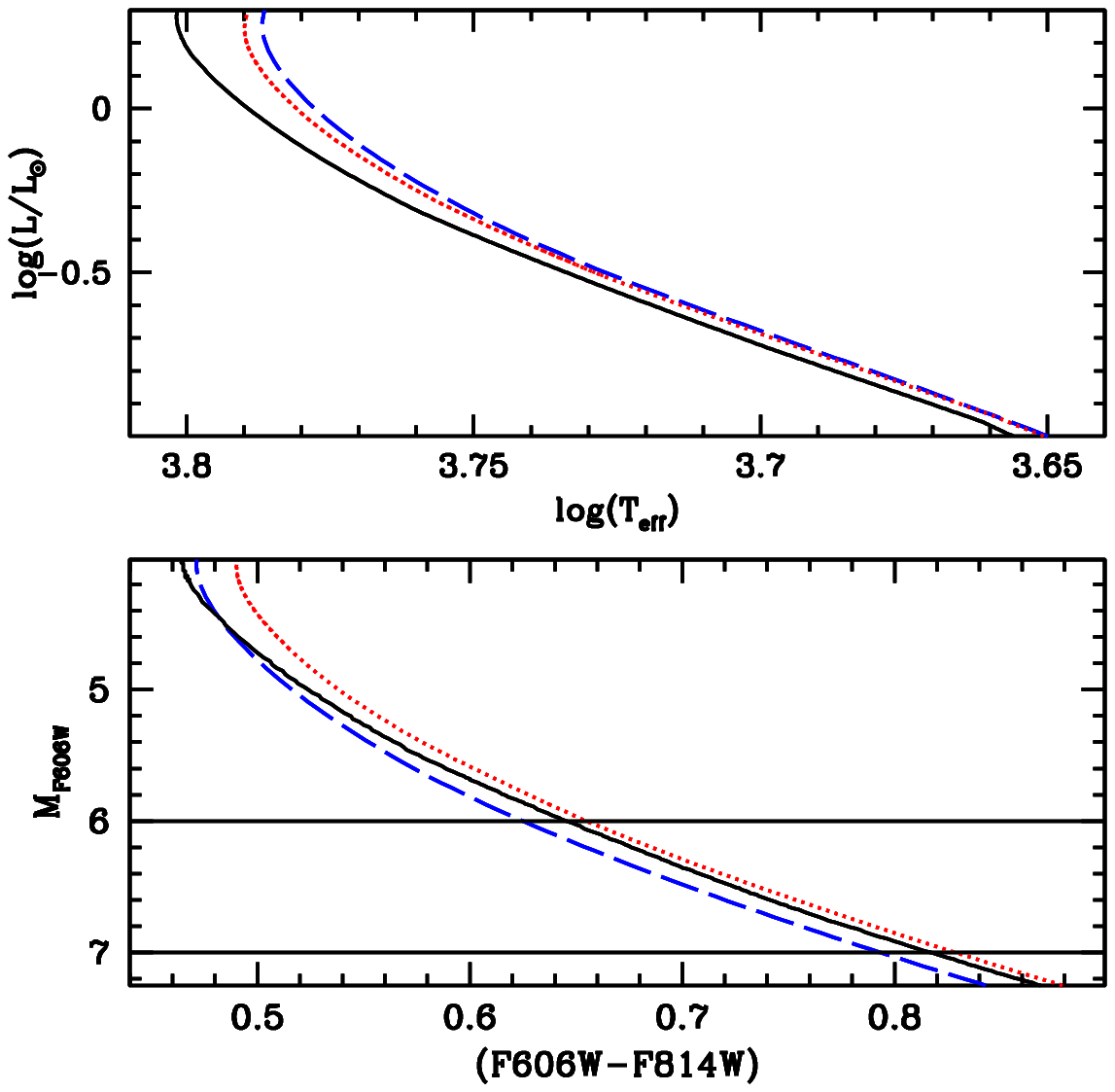}
\caption{Comparison of \citet{pisa} and  \citet{vdbmodels} 12~Gyr, $Z$=0.002, $Y$=0.25 isochrones 
(dotted and dashed lines, respectively) with our reference 12~Gyr, $Z$=0.002, $Y$=0.248 
isochrone (solid line), in both the $L$-$T_{eff}$ (upper panel) and $F606W-(F606W-F814W)$ (lower panel) 
diagrams}
\label{isoauthors}
\end{figure}

Figure~\ref{isoauthors} compares the $Y$=0.25 isochrone (the only one with initial $Y$ very close to 
our adopted reference values) to our reference one with $Y$=0.248. At fixed $L$ the \citet{pisa} isochrone is slightly cooler 
along the unevolved MS, by about $\sim$70~K. Differences increase when moving towards the turn off, because of the inclusion of fully efficient diffusion 
in \citet{pisa} calculations. The same behaviour appears obviously (the same 
BCs are used) when considering colour differences at fixed $F606W$ magnitude. Along the 
unevolved MS \citet{pisa} isochrones are redder by less than 0.01~mag.

Considering a reference magnitude $M_{F606W}$=7, where the effect of atomic diffusion is negligible, 
we derive d$Y$/d($F606W-F814W$)=2.0 for the \citet{pisa} isochrones, that traslates to an increase of just 10\% 
in the derived $\Delta Y$ values, compared to our reference calibration. 

The \citet{vdbmodels} stellar models are calculated with [$\alpha$/Fe]=0.4 and the reference solar metal mixture by \citet{agss}. 
They include the gravitational settling of helium, Li and Be but not of heavy elements. 
The efficiency of settling from the convective envelope is then moderated as described in \citet{vdbphys}. 
Equation of state and reaction rates for hydrogen burning are also different from our reference calculations \citep[see][for details]{vdbphys}. 
The outer boundary conditions for the masses relevant to our analysis are obtained integrating the empirical solar $T(\tau)$ relation 
by \citet{hm}, also different from our reference choice.
The adopted BCs are the MARCS BCs by \citet{cv} we have already discussed.

We considered their 12~Gyr, $Z$=0.002, $Y$=0.25, 0.29 and 0.33 isochrones to calculate d$Y$/d($F606W-F814W$), and plotted in Fig.~\ref{isoauthors} the $Y$=0.25 one. 
At fixed $L$ the \citet{vdbmodels} is essentially identical to the \citet{pisa} counterpart, that is slightly cooler than our reference isochrone 
along the unevolved MS. Moving towards the turn off the \citet{vdbmodels} isochrone becomes redder than the \citet{pisa} counterpart, 
most probably because of the different implementation of atomic diffusion.
The behaviour in the $F606W-(F606W-F814W)$ diagram is however different. Because MARCS BCs produce systematically bluer colours 
(at fixed $M_{F606W}$) compared to the ATLAS9 BCs (see Fig.~\ref{isobcs}), the \citet{vdbmodels} isochrone appears now bluer than the 
our reference one (apart from the turn off region), by about 0.02~mag.
Despite this colour difference at fixed $Y$, we find that the derivative d$Y$/d($F606W-F814W$) (taken at 
$M_{F606W}$=7, where the effect of atomic diffusion is negligible) differs by less than 5\%, compared 
to the result with our reference isochrones.

\section{Summary}
\label{conclusions}

Estimates of the He abundance range $\Delta Y$ within individual GGCs are crucial to constrain the origin of the multiple population phenomenon, 
and the use of the MS width in optical CMDs is, from the theoretical point, of view a very reliable reliable approach to constrain this quantity.
The availability of both UV (from the $HST$-WFC3 camera) and optical (from the $HST$-ACS camera) magnitudes 
thanks to the {\sl Hubble Space Telescope UV Legacy Survey of Galactic GCs} project, will allow 
the homogeneous determination of $\Delta Y$ for a large sample of GGCs. The UV CMDs can efficiently disentangle the various sub-populations, 
and MS colour differences in the ACS  $F606W-(F606W-F814W)$ CMD will allow an estimate of their initial He abundance differences. 

We have established here that the d$Y$/d$(F606W-F814W)$ colour derivative at fixed $M_{F606W}$ (or $M_{F814W}$) between $\sim$2 and $\sim$3 magnitudes below the 
MS turn off is weakly affected by 
current uncertainties in a range of physics inputs that enter the calculations of stellar models 
and isochrones, and is completely unaffected by the presence of CNONa abundance anticorrelations. 

The value of d$Y$/d$(F606W-F814W)$ (between ) generally decreases with increasing [Fe/H],  
and, to give a quantitative example, changes by $\sim$25\% at $M_{F606W}=7$ when going from [Fe/H]=$-$1.6 to $-$0.7. 
Table~\ref{table} summarizes the results for [Fe/H] ranging between $-$2.1 and $-$0.7.

\begin{table}
\caption{Derivatives d$Y$/d($F606W-F814W$) for our reference calculations, taken at $\sim$2 and $\sim$3~mag below the $F606W$ turn off 
magnitude for an age of 12~Gyr}
\label{symbols}
\begin{tabular}{lcc}
\hline
[Fe/H] & d$Y$/d($F606W-F814W$) & d$Y$/d($F606W-F814W$)\\
  &  MS turn off +1 mag  & MS turn off + 2 mag\\
\hline
$-$2.1 & 2.3 & 2.1 \\
$-$1.6 & 2.2 & 1.9 \\
$-$1.3 & 2.1 & 1.8 \\
$-$1.0 & 2.0 & 1.6 \\
$-$0.7 & 1.7 & 1.5 \\
\hline
\label{table}
\end{tabular}
\end{table}

We have found that, at our selected [Fe/H]=$-$1.3 taken as representative of the results that we obtain over all GGC metallicity range, 
variations by $\pm$0.1 of $\alpha_{\rm MLT}$ 
around the solar calibrated value, and the use of different $T(\tau)$ relationships to determine the outer boundary conditions for the model calculations,  
leave d$Y$/d$(F606W-F814W)$ unchanged, although they change slightly the colours of the isochrones, at the level of at most $\sim$0.02~mag.

Uncertainties due to adopted BCs are not large, in the assumption that our reference ATLAS9, PHOENIX, MARCS and \citet{wl} BCs we employed in these 
comparisons 
are a fair reflection of current uncertainties. The \citet{wl} BCs induce a variation of d$Y$/d($F606W-F814W$) compared to our reference result that causes 
a decrease of $\Delta Y$ by 5\% if estimated at $M_{F606W}$=6, and by 16\% if estimated at $M_{F606W}$=7. The PHOENIX and MARCS 
BCs do not change d$Y$/d($F606W-F814W$).

Just to give an idea of the corresponding variations of the absolute values of $\Delta Y$, current estimates provide a typical upper limit $\Delta Y\sim$0.05 
\citep[apart from exceptions like NGC2808 and $\omega$~Centauri, see, e.g., the data compiled in Table 1 of][]{bastian:15}. A 20\% 
increase or decrease corresponds to variations by at most $\sim$0.01.

If atomic diffusion is fully efficient throughout the star, 
the value of d$Y$/d($F606W-F814W$) measured at $M_{F606W}=6$ is affected at the level of at most $\sim$20\%, if $Y$ is between 0.35 and 0.40.
However, spectroscopic observations of GGC stars have shown that diffusion from the convective envelopes of GGC stars is severely inhibited, and 
when we artificially stop diffusion from the envelope, keeping it efficient only in the underlying radiative layers, 
we find that d$Y$/d($F606W-F814W$) is unaffected compared to the no-diffusion case.
  
Finally, the independent isochrone libraries of \citet{pisa} amd \citet{vdbmodels} provide a d$Y$/d($F606W-F814W$) (in the 
magnitude range where atomic diffusion is negligible) only at most 10\% higher than our reference set, despite 
various differences in the model physics inputs involving the equation of state, outer boundary conditions, nuclear reaction rates 
and metal abundance mixture.

In summary, we found that the theoretical framework to estimate He variations amongst sub-populations 
in individual GGCs through MS colour differences is quite robust. Although altering several of the required physics 
inputs/assumptions can affect the MS colours for a given initial chemical composition, 
colour differences due to variations of the initial $Y$ abundances are a robust prediction of the current generation of stellar models.

\section*{acknowledgements}

We thank the anonymous referee for his/her report that has greatly helped to improve the presentation of our results. 
SC acknowledges the financial support by PRIN-INAF2014 (PI: S. Cassisi), and the Economy and Competitiveness Ministry of the 
Kingdom of Spain (grant AYA2013-42781-P). AP warmly thanks for financial support by PRIN-INAF2012 (PI: L. Bedin).

\bibliographystyle{mnras}


\bsp	
\label{lastpage}
\end{document}